\documentclass[fleqn]{iopart}
\usepackage{iopams}
\usepackage{amsfonts,amssymb,amsthm}
\usepackage{latexsym}
\usepackage{bbm,bm}
\usepackage{graphicx}
\usepackage{xcolor}
\def\mbf{\mathbf}
\def\bm{\boldmath}
\def\mcl{\mathcal}
\def\dst{\displaystyle}

\def\beq{\begin{eqnarray*}}
\def\eeq{\end{eqnarray*}}

\def\bs{\boldsymbol}

\begin{document}

\title{Aharonov-Bohm-Coulomb Problem in Graphene Ring}

\author{Eylee Jung$^1$, Mi-Ra Hwang$^2$, Chang-Soo Park$^3$ and DaeKil Park$^{1,2}$}

\address{$^1$Department of Physics, Kyungnam University, ChangWon, 631-701, Republic of Korea}
\address{$^2$Department of Electronic Engineering, Kyungnam University, ChangWon, 631-701, Republic of Korea}
\address{$^3$Department of Physics, Dankook University, Cheonan 330-714, Republic of Korea}
\eads{\mailto{dkpark@kyungnam.ac.kr}, \mailto{olnal@dankook.ac.kr}}
\date{\today}

\begin{abstract}
We study the Aharonov-Bohm-Coulomb problem in a graphene ring. We investigate, in particular, the effects of a Coulomb type potential of the form $\xi/r$ on the energy spectrum of
Dirac electrons in the graphene ring in two different ways: one for the scalar coupling and the other for the vector coupling. It is found that, since the  potential in the scalar
coupling breaks the time-reversal symmetry between the two valleys as well as the effective time-reversal symmetry in a single valley, the energy spectrum of one valley is
separated from that of the other valley, demonstrating a valley polarization. In the vector coupling, however, the potential does not break either of the two symmetries and its
effect appears only as an additive constant to the spectrum of Aharonov-Bohm potential. The corresponding persistent currents, the observable quantities of the symmetry-breaking energy
spectra, are shown to be asymmetric about zero magnetic flux in the scalar coupling, while symmetric in the vector coupling.
\end{abstract}
\submitto{\JPA}
\maketitle

%---------------------------------------------------------------------------
\section{Introduction}
\begin{figure}[h]
\begin{center}
\includegraphics[height=4cm]{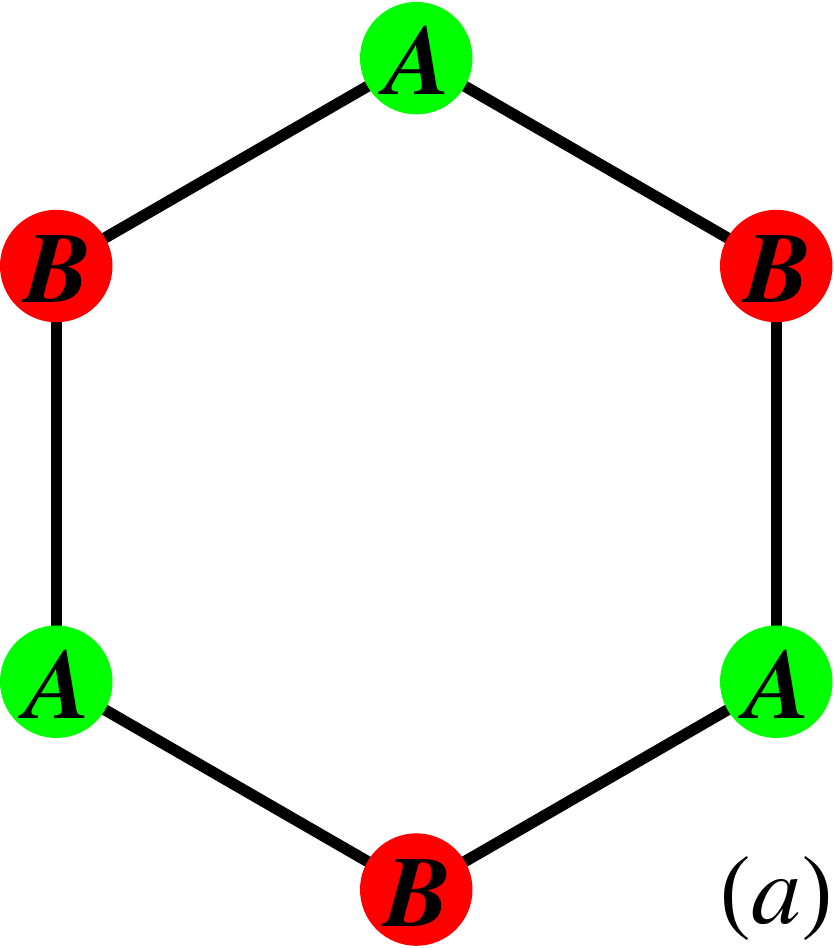}
\hspace{0.7in}
\includegraphics[height=3.6cm]{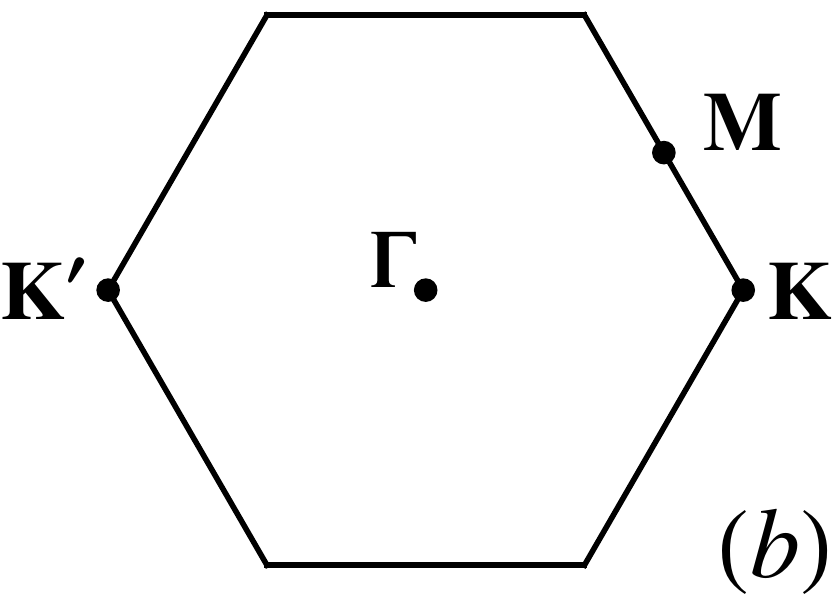}
\end{center}
\caption{\label{fig1}The Honeycomb lattice (left) consists of two sublattices, one with $A$ atoms and the other with $B$ atoms. The right is the corresponding  Brillouin zone and
its symmetry points. Note that the two symmetry points, $\bs{K}$ and $\bs{K'}$, are related by $\bs{K'} = -\bs{K}$. The energy spectra near these points for conduction band are
given by $E =\hbar v_F |\bs{k}\pm \bs{K}|$, which are the conic dispersion relations, called valleys.}
\end{figure}

Graphene is a two-dimensional composite lattice with honeycomb structure and consists of two sublattices of carbon atoms. The corresponding Brillouin zone (BZ) is also a hexagon
with high symmetry points at vertices as well as center and side (see Fig.~\ref{fig1}) \cite{wallace,neto09}. Near the vertices of the BZ the low-energy electronic spectra are
linearly dependent on the magnitude of momentum, forming conical valleys, and the dynamics of electrons in graphene can be formulated by Dirac equation of massless fermion
\cite{slonczewski,semenoff84}. Because of this Dirac fermion-like behavior of electrons, graphene may offer an opportunity to test the various predictions of the planar field
theories \cite{haldane,jackiw90} by experiment with solid state material. In this respect there have been much effort to connect the two-dimensional (2D) field theory with the
graphene physics \cite{connection}, including the Aharonov-Bohm (AB) effect~\cite{ab1959} in a ring geometry of graphene~\cite{recher07, wurm, ab-graphene-2, ab-graphene-exp}.

The studies of the AB effect in a graphene ring were mostly concerned about the breaking of time-reversal symmetry (TRS) that yields a splitting of the valley degeneracy due to
the time-reversal symmetry in graphene.  In this regard, the authors of Ref.~\cite{recher07} have demonstrated that the splitting of the degeneracy in a single valley can be
controlled by the threaded magnetic flux and a confinement potential of the Dirac electrons on the AB ring of graphene. An interesting result from the work was that the
confinement potential, introduced as a mass term in the Dirac equation, leads to a breaking of the TRS in the absence of the magnetic flux.

In this paper, motivated from the above result, we study the Aharonov-Bohm-Coulomb (ABC) problem~\cite{kibler,villalba} in a graphene ring to investigate the effect of a Coulomb
type potential in the form of $\xi/r$ on the splitting of valley degeneracy.~\footnote{Thus it includes both the electric Coulomb potential and a relativistic scalar potential.}
Within the framework of field theory there are many possible ways in introducing the potential to the Dirac equation. The only criterion for a consistent coupling of the potential
to the Dirac equation is the fact that the larger component of the Dirac field $\varphi$ should satisfy the Schr\"{o}dinger equation $\left[p^2/2M +\xi/r\right]\varphi =
\mcl{E}\varphi$ in the non-relativistic Galilean limit \cite{galilean}. In fact, there are at least two ways such that the same consistent Galilean limit is fulfilled. In the
present work we consider the following two possibilities~\cite{coupling}: one is the \emph{scalar coupling} and the other is the \emph{vector coupling}. As we shall describe
explicitly in next section, in the former case the $\xi/r$ potential enters the Dirac equation as a mass term, while it enters the equation as an energy term in the latter case.
From the viewpoint of time-reversal symmetry the scalar coupling is expected to break the TRS, but the vector coupling is not. In the following, we explicitly show that the scalar
coupling indeed leads to the splitting of valley degeneracy, while the vector coupling does not. What is remarkable and new in our result is that, besides the splitting of the
single-valley degeneracy, the scalar coupling produces a separation between the energy spectrum of one valley and that of the other valley, and the separation increases with the
interaction strength $\xi$.

The paper is organized as follows. In Sec.~\ref{sec2}, starting with a brief recapitulation of the time-reversal symmetries in graphene, we give a qualitative argument
how the TRS is broken by the $\xi/r$ potential in the AB ring of graphene, whereby the splitting of valley degeneracy is produced. In Sec.~\ref{sec3} we solve the 2D Dirac
equation for the scalar coupling of the potential in a graphene ring. We derive an analytical expression of the energy spectrum in terms of valley parameter (denoted by $\tau$)
and the interaction strength $\xi$. These two parameters interplay to separate the whole energy spectrum of one valley from the other. We then compute a persistent current in the
ring. Owing to the separation of energy spectra the persistent current is asymmetric about the zero magnetic flux, which represents an essentially single-valley characteristic in
the case of scalar coupling, known as the valley polarization~\cite{akhmerov, rycerz}. In Sec.~\ref{sec4}, we present the energy spectrum for the vector coupling. It is shown
that, contrary to the scalar coupling case, the potential via vector coupling to the Dirac equation alone can break neither the intravalley nor the intervalley degeneracies
because of the conservation of the TRS. We find that the potential in the vector coupling serves effectively as an additive constant in the energy spectrum and is completely
decoupled from the AB effect. Finally, there will be a conclusion in Sec.~\ref{sec5}.

%%%%%%%%%%%%%%%%%%%%%%%%%%%%%%%%%%%%%%%%%%%%%%%%%%%%%%%%%%%%%%%%%%%%%%%%%%%%%%%%%%%%%%%%%%%%%%%%%%%%%%%%
%%%%%%%%%%%%%%%%%%%%%%%%%%%%%%%%%%%%%%%%%%%%%%%%%%%%%%%%%%%%%%%%%%%%%%%%%%%%%%%%%%%%%%%%%%%%%%%%%%%%%%%%%%%%

\section{TRS breaking and splitting of valley degeneracy}\label{sec2}
In this section we first briefly recapitulate the discrete symmetry of the time reversal in graphene, then discuss the symmetry breaking in the ABC problem of graphene ring. As
illustrated in Fig.~\ref{fig1} there are two inequivalent symmetry points $\bs{K}$ and $\bs{K'}$ in the BZ of a graphene, and to each point two sublattices $A$ and $B$ are
associated. The Hamiltonian of a Dirac electron in graphene is conveniently described by two pseudospins related to the two valleys at $\bs{K}$ and $\bs{K'}$ and the two
sublattices. In the following we use the Pauli matrices $\bs{\tau} = (\tau_1,\tau_2,\tau_3)$ and $\bs{\sigma} = (\sigma_1,\sigma_2,\sigma_3)$ to denote the valley and sublattice
degree of freedom, respectively and also use the $2\times2$ unit matrices $\tau_0$ and $\sigma_0$. Based on the geometry in Fig.~\ref{fig1} the Hamiltonian and corresponding
four-component spinor can be described as
\begin{eqnarray}\label{ham1}
\dst\fl H_0 = \tau_0\otimes H_K,\quad H_K =  v_F\bs{\sigma}\cdot\mbf{p},\quad \Psi = (\Psi_{AK},\Psi_{BK},-\Psi_{BK'},\Psi_{AK'})^T,
\end{eqnarray}
where $v_F\sim c/300$ is the Fermi velocity, $T$ stands for transpose, and we have used the \emph{valley-isotropic} form of hamiltonian for convenience of subsequent
calculations~\cite{akhmerov}.

To see the TRS in graphene we introduce the following time-reversal operator
\begin{eqnarray}\label{trs1}
\dst \eqalign{\mcl{T} = \left[(i\tau_2)\otimes(i\sigma_2)\right]\mcl{C},}
\end{eqnarray}
where $\mcl{C}$ is the complex conjugate operator. The effect of $\mcl{T}$ on the Hamiltonian $H_0$ and the state $\Psi$ are
\begin{eqnarray}\label{trsproof1}
\dst \eqalign{\mcl{T}H_0\mcl{T}^{-1} = H_0,\quad \mcl{T}\Psi = (\Psi_{AK'}^*,\Psi_{BK'}^*,-\Psi_{BK}^*,\Psi_{AK}^*)^T.}
\end{eqnarray}
This shows that the Hamiltonian $H_0$ is invariant under the transformation by $\mcl{T}$ but it interchanges the valleys; there exists a degeneracy between the two valleys
(henceforth, intervalley degeneracy). According to Ref.~\cite{ando02}, the Hamiltonian satisfies another TRS under the transformation by the operator
\begin{eqnarray}\label{trs2}
\dst \eqalign{\bs{\Theta} = \tau_0\otimes \bs{\Theta}_{\sigma},\quad \bs{\Theta}_{\sigma} = i\sigma_2\mcl{C}.}
\end{eqnarray}
With this operator one can verify
\begin{eqnarray}\label{trsproof2}
\dst \eqalign{\bs{\Theta}H_0{\bs{\Theta}}^{-1} = H_0,\quad \bs{\Theta}\Psi = (\Psi_{BK}^*,-\Psi_{AK}^*,\Psi_{AK'}^*, \Psi_{BK'}^*)^T.}
\end{eqnarray}
This operator exchanges the sublattices within a single valley, but does not interchange the valleys as we can see in the right equation. Since the Hamiltonian $H_0$ is invariant
this leads to a degeneracy within a single valley (henceforth, intravalley degeneracy)~\cite{beenakker08}. More specifically, the operator $\bs{\Theta}_{\sigma}$ transforms $H_K$
in Eq.~(\ref{ham1}) to $\bs{\Theta}_{\sigma}H_K\bs{\Theta}_{\sigma}^{-1} = H_K$, that is, $H_K$ is invariant as the intravalley degeneracy implies: $\bs{\Theta}_{\sigma}$
effectively changes the sign of $\bs{\sigma}$ and $\mbf{p}$.

Having introduced the TRS's of graphene we now discuss the TRS breaking of the ABC problem in a graphene ring. For this we first consider an AB type ring of a graphene discussed
in Ref.~\cite{recher07}: inner and outer radii of the ring are $a + w/2$ and $a-w/2$, so that the ring width is $w$.  Since the Hamiltonian $H_0$ has a valley-isotropic form it is
more convenient to use $H_K$ with two-component spinor and insert the valley index in an appropriate place. For the AB ring the Hamiltonian then reads~\footnote{Throughout this
paper we use the convention $\hbar=c=1$ unless otherwise specified.}
\begin{eqnarray}\label{ham2}
\dst H(\bs{A}) = H_K(\bs{A}) +  \tau V(r)\sigma_3,\quad H_K(\bs{A}) = v_F\bs{\sigma}\cdot \left(\bs{p} +e \bs{A}\right)
\end{eqnarray}
where the circularly symmetric potential $\tau V(r)\sigma_3$ is introduced to confine a Dirac electron on the graphene ring: $V(r)=0$ when $|r-a| \leq w/2$ and
$V(r)\rightarrow\infty$ when $|r-a| \geq w/2$. The index $\tau=\pm 1$ denotes the valleys: $\tau=+1$ for the $\bs{K}$ valley and $\tau=-1$ for the $\bs{K'}$ valley. We note here
that this potential is proportional to $\sigma_3$.

Obviously the magnetic field breaks the effective TRS of $\bs{\Theta}_{\sigma}$: ${\bs\Theta}_{\sigma}H_K({\bs A}){\bs\Theta}_{\sigma}^{-1} = H_K(-{\bs A})$. In fact, the TRS of
the Hamiltonian with four-component spinor is also broken: $\mcl{T}[\tau_0\otimes H_K({\bs A})]\mcl{T}^{-1} = \tau_0\otimes H_K({-\bs A})$. Thus, the presence of magnetic field
will lift both of the intervalley and intravalley degeneracies. What is interesting in Eq.~(\ref{ham2})  is the TRS breaking by the confinement potential term when $\bs{A} = 0$.
In the absence of magnetic field one can see
\begin{eqnarray}
\dst \eqalign{\bs{\Theta}_{\sigma}[H_K(0) +  \tau V(r)\sigma_3]\bs{\Theta}_{\sigma}^{-1} = H_K(0) -  \tau V(r)\sigma_3.}
\end{eqnarray}
This implies the intervalley degeneracy can be broken by the confinement potential as demonstrated in Ref.~\cite{recher07}. According to Berry~\cite{berry87} this potential enters
the 2-D Dirac equation by the replacement of $Mc^2\rightarrow Mc^2 + V(r)\sigma_3$ with $M=0$ for a massless particle. In this sense the confinement potential is regarded as a
mass term in the Dirac equation, and it can be conjectured that a mass term proportional to $\sigma_3$ leads to a breaking of the effective TRS in an AB graphene ring.

We now turn to the problem of ABC in a graphene ring. Here we need to include a Coulomb type potential $\xi/r$ in the Dirac equation. To address the TRS breaking by this
potential we begin with the general form of a 2D relativistic Dirac equation minimally coupled to AB potential $\bs{A}$, which can be written as
\begin{equation}\label{dirac-1}
\dst \eqalign{\left[c\beta {\bm \gamma} \cdot {\bm \Pi} + \beta Mc^2 \right] \psi = E \psi,}
\end{equation}
where $M$ is the bare mass of a Dirac electron, $\psi$ is the two-component spinor, and $\Pi_i = -i \partial_i - e A_i$ is the covariant derivative multiplied by $-i$. The Dirac
matrix is chosen as
\begin{eqnarray}
\label{dirac-matrix}  \eqalign{\beta = \sigma_3,\quad \beta \gamma_i = (\sigma_1, s \sigma_2),\quad \{\gamma^{\mu}, \gamma^{\nu} \} = 2 \eta^{\mu \nu},\quad \eta^{\mu \nu} = (+,
-, -),}
\end{eqnarray}
where $s$ is twice of the spin value ($+1$ for spin ``up'' and $-1$ for spin ``down''). We employ a thin flux tube as the AB potential in the form
\begin{equation}
\label{ab-potential} e A_i = \frac{\alpha \epsilon_{ij} r_j}{r^2},
\end{equation}
which yields the magnetic field $B = -2 \pi \alpha \delta({\bm r}) / e$ along the $z$-direction as expected. Therefore, the parameter $\alpha$ represents a magnetic flux in unit
of $-e/ (2 \pi)$.

As mentioned in introduction there are at least two possibilities for the potential $\xi/r$~\footnote{It should be noted that this is a three-dimensional expression of
the Coulomb potential. The true 2D expression of the Coulomb potential is proportional to $\ln r$. The reason for use of $\xi/r$ is due to the fact 
that the two-dimensional graphene is embedded in the three-dimensional space.} to be included in the Dirac equation, the scalar coupling and the vector coupling. For the scalar
coupling it couples to the equation as
\begin{eqnarray}
\dst \eqalign{\left[c\beta {\bm \gamma} \cdot {\bm \Pi} + \beta\left(Mc^2 + \frac{\xi}{r}\right)\right] \psi = E \psi,}
\end{eqnarray}
and for the vector coupling the equation becomes
\begin{eqnarray}
\dst \eqalign{\left[c\beta {\bm \gamma} \cdot {\bm \Pi} + \beta Mc^2\right] \psi = \left(E - \frac{\xi}{r}\right)\psi.}
\end{eqnarray}
From a physical point of view the potential $\xi/r$ in the vector coupling corresponds to the electric Coulomb potential 
(time component of the relativistic four vector) and 
the potential in the scalar coupling is a relativistic scalar potential (other than the
electric Coulomb potential). For the Dirac electron in a graphene, since $M = 0$ and $c = v_F$, we have
\begin{eqnarray}
\dst && H_s\psi = E\psi,\quad H_s(\bs{A}) =   H_K(\bs{A}) + \frac{\xi}{r}\sigma_3,\label{hscalar} \\
\dst && H_v\psi = E\psi,\quad H_v(\bs{A}) =   H_K(\bs{A}) + \frac{\xi}{r},\label{hvector}
\end{eqnarray}
where $H_s(\bs{A})$ and $H_v(\bs{A})$ stand for Hamiltonians of scalar coupling and vector coupling, respectively, $H_K(\bs{A})$ is given in Eq.~(\ref{ham2}), and we have used the
conventions in Eq.~(\ref{dirac-matrix}) with $s=+1$.

As before the magnetic field will break both of the TRS's for $\mcl{T}$ and $\bs{\Theta}$. To see the effect of the  potential $\xi/r$, we take the time-reversal transformation on
the Hamiltonians for $\bs{A}=0$. For the operator $\mcl{T}$ we get
\begin{eqnarray}
\dst  \eqalign{\mcl{T}\tau_0 \otimes H_s(0)\mcl{T}^{-1}  = \tau_0\otimes \left[H_K(0) - \frac{\xi}{r}\sigma_3\right], \cr \dst  \mcl{T}\tau_0 \otimes H_v(0)\mcl{T}^{-1}  =
\tau_0\otimes \left[H_K(0) + \frac{\xi}{r}\right].}
\end{eqnarray}
The scalar coupling breaks the TRS for $\mcl{T}$, but the vector coupling preserves the TRS. For the effective TRS operator $\bs{\Theta}_{\sigma}$ the two Hamiltonians  are
transformed to
\begin{eqnarray}
\dst \eqalign{ \bs{\Theta}_{\sigma}H_s(0)\bs{\Theta}_{\sigma}^{-1}  = H_K(0) - \frac{\xi}{r}\sigma_3, \cr \dst  \bs{\Theta}_{\sigma} H_v(0)\bs{\Theta}_{\sigma}^{-1} =  H_K(0) +
\frac{\xi}{r}.}
\end{eqnarray}
Here we have the same results: the scalar coupling breaks the effective TRS for $\bs{\Theta}_{\sigma}$, but the vector coupling does not. Therefore we expect that the scalar
coupling will lift both of the intervalley and intravalley degeneracies. On the other hand, the vector coupling breaks neither of the two degeneracies. In the following sections
we will compute the energy spectra of the ABC problem in a graphene ring to show the effects of each coupling explicitly.

\section{ABC problem with scalar coupling}\label{sec3}
In this section we compute the energy spectrum for the scalar coupling of ABC in a graphene ring. As described before the geometry of the ring is the same as discussed
in Ref.~\cite{recher07}. Using the Hamiltonian (\ref{hscalar}) and the confinement potential in Eq.~(\ref{ham2}) the 2D Dirac equation is
\begin{equation}\label{scalar-hamiltonian-1}
\dst H_s\psi = E\psi,\quad  H_s = \left[v_F \bs{\sigma} \cdot \left(\bs{p} + e\bs{A}\right) + \frac{\xi}{r}\sigma_3\right] + \tau V(r) \sigma_3,
\end{equation}
where $\tau=\pm$ is the valley index. As defined earlier, in the confinement potential, $V(r) = 0$ when $|r-a| \leq w/2$ and $V(r) \rightarrow \infty$ when $|r-a|
>w/2$. According to Refs.~\cite{berry87} and \cite{mccann} the boundary conditions on the two-component spinor $\psi$ can be expressed as the form
\begin{eqnarray}\label{boundary}
\dst \psi = \tau({\bs n}_{\perp}\cdot {\bs \sigma})\psi,\quad \left\{\begin{array}{ll} {\bs n}_{\perp} = (-\sin\theta,\, \cos\theta), &~ r=a+\frac{w}{2}\\ {\bs n}_{\perp} =
-(-\sin\theta,\, \cos\theta), &~ r=a-\frac{w}{2} \end{array}\right.
\end{eqnarray}
where ${\bs n}_{\perp}$ is the unit vector perpendicular to the normal direction ${\bs n}$ on the boundary (i.e, ${\bs n}_{\perp}\cdot{\bs n} = 0$) and lies in $(x,y)$ plane, that
is, on the ring plane. This choice of boundary conditions is based on the requirement of the hermiticity of the Hamiltonian $H$ within the boundary, which yields the condition of
no outward current at any point on the ring boundary (i.e., ${\bs n}\cdot {\bs v}= 0$, where ${\bs v} = <\psi| v_F{\bs \sigma} |\psi>$).~\footnote{More precisely, the
boundary conditions are determined by the \emph{self-adjointness} of the Hamiltonian $H$ in (\ref{scalar-hamiltonian-1}): $<\psi|H\psi> - <H\psi|\psi> = 0$. Mathematically, this
is related to the self-adjoint extension with deficiency indcies $(2,2)$, so that the two-component spinor satisfies $\psi = U\psi$ at boundaries, where $U$ is a $2\times2$
unitary, hermitian matrix with unit determinant ~\cite{capri,guy}. In the present case $U = {\bs n}_{\perp}\cdot {\bs \sigma}$ with which the operator ${\bs \Theta}_{\sigma} =
i\sigma_2\mcl{C}$ anticommutes, that is, $\{U,{\bs \Theta}_{\sigma}\} = 0$. Thus, the chosen boundary condition does not preserve the effective TRS.
This particular choice of $U$ is to prevent the Klein tunneling.}

For a Dirac electron inside the ring the Dirac equation reads, using $V(r)=0$,
\begin{equation}
\label{dirac-scalar-1} \left[{\bs\sigma} \cdot \left(\bs{p} + e\bs{A}\right) + \frac{\tilde{\xi}}{r}\sigma_3\right]\psi = \tilde{E} \psi
\end{equation}
where $\tilde{E} = E / v_F$ and $\tilde{\xi} = \xi/v_F$. Operating $[-{\bs\sigma} \cdot \left(\bs{p} + e\bs{A}\right) + \sigma_3\tilde{\xi}/r + \tilde{E} ]$ on the equation and
using polar coordinates we have second order equations
\begin{eqnarray}
\label{second-scalar-1} \eqalign{ \left[ \partial_r^2 + \frac{1}{r} \partial_r + \frac{1}{r^2} \left( \partial_{\theta} + i \alpha \right)^2 - \frac{\tilde{\xi}^2}{r^2} +
\tilde{E}^2 + s e B \sigma_3 \right] \psi \cr = \frac{i \tilde{\xi}}{r^2} \left(     \begin{array}{cc}
                                                        0  &  -e^{-i s \theta}       \\
                                                        e^{i s \theta}  &  0
                                                            \end{array}             \right)    \psi,}
\end{eqnarray}
where $B$ and $\alpha$ are the magnetic field and flux given in Eq.~(\ref{ab-potential}).  Since the Hamiltonian (\ref{scalar-hamiltonian-1}) satisfies $[H, J_3] = 0$, where $J_3
= -i
\partial_{\theta} + (s/2) \sigma_3$, the solution to the Dirac equation can be written in the form
\begin{eqnarray}
\label{solution-scalar-1} \psi(r,\theta) = \left(       \begin{array}{c}
              \chi_{1 m} (r) e^{i (m - s/2) \theta}       \\
              \chi_{2 m} (r) e^{i (m + s/2) \theta}
                     \end{array}                       \right),\quad\left(m=\pm \frac12, \pm\frac32, \cdots\right).
\end{eqnarray}
Inserting this into Eq.~(\ref{second-scalar-1}) we can extract the radial equation
\begin{eqnarray}
\label{radial-1} \eqalign{\left[ \partial_r^2 + \frac{1}{r} \partial_r - \frac{(m + \alpha)^2 + \tilde{\xi}^2 + 1/4}{r^2} + \tilde{E}^2 \right] \left(
\begin{array}{c}\chi_{1m}(r)\\  \chi_{2m}(r)
\end{array}\right)\cr = -\frac{1}{r^2} (\eta \sigma_3 - \tilde{\xi} \sigma_2) \left(\begin{array}{c} \chi_{1m}(r)\\  \chi_{2 m}(r)\end{array}\right)},
\end{eqnarray}
where
\begin{equation}
\label{eta-def-1} \eta = m + \alpha.
\end{equation}
Using the matrix diagonalization the right hand side can be written as
\begin{eqnarray}
\label{diagonal-1} \dst \eqalign{\eta \sigma_3 - \tilde{\xi} \sigma_2 = \epsilon(\eta) \sqrt{\eta^2 + \tilde{\xi}^2}\, U_c^{\dagger} \sigma_3 U_c     \cr U_c = \cos
\frac{\phi}{2}\sigma_0 + i \sin \frac{\phi}{2} \sigma_1 \quad  \left(-\frac{\pi}{2} \leq \phi = \tan^{-1} \frac{\tilde{\xi}}{\eta} \leq \frac{\pi}{2}\right)},
\end{eqnarray}
where $\epsilon(x) = |x| / x$ is the alternating function, $\sigma_0$ is the $2\times2$ unit matrix, and
\begin{eqnarray}
\label{tan-phi-2} \dst & &\tan \frac{\phi}{2} = \epsilon(\eta) \frac{\sqrt{\eta^2 + \tilde{\xi}^2} - |\eta|}{\tilde{\xi}}.
\end{eqnarray}
With this diagonalization the radial equation (\ref{radial-1}) reduces to
\begin{eqnarray}
\label{radial-2} \eqalign{\left[ \partial_r^2 + \frac{1}{r} \partial_r - \frac{(m + \alpha)^2 + \tilde{\xi}^2 + 1/4}{r^2} + \tilde{E}^2 \right] \left( \begin{array}{c} f_{1
m}(r)\\ f_{2 m}(r)
\end{array}\right)\cr = -\frac{1}{r^2} \epsilon(\eta) \sqrt{\eta^2 + \tilde{\xi}^2}\, \sigma_3 \left( \begin{array}{c} f_{1 m}(r)\\ f_{2 m}(r)\end{array}\right),}
\end{eqnarray}
where the two components $f_{1m}(r)$ and  $f_{2m}(r)$ are related to the spinor $\chi(r)$ as
\begin{eqnarray}
\dst \chi(r) = \label{def-f} \left( \begin{array}{c} \chi_{1 m}(r)\\  \chi_{2 m}(r) \end{array}\right) = U_c^{\dagger} \left( \begin{array}{c} f_{1 m}(r)\\ f_{2 m}(r) \end{array}\right).
\end{eqnarray}

The reduced equation for each component in Eq.~(\ref{radial-2}) is the Bessel's equation and, introducing dimensionless radial variable $\rho = |\tilde{E}|r$, the solutions can be expressed as
\begin{eqnarray}
\label{solution-1} \eqalign{f_{1 m}(\rho) = A_{1 m} H_{\nu_-}^{(1)} (\rho) + B_{1 m} H_{\nu_-}^{(2)} (\rho)    \cr  f_{2 m}(\rho) = A_{2 m} H_{\nu_+}^{(1)} (\rho) + B_{2 m}
H_{\nu_+}^{(2)} (\rho),}
\end{eqnarray}
where $H_{\nu}^{(1)} (\rho)$ and $H_{\nu}^{(2)} (\rho)$ are the Hankel functions, and the orders are given by
\begin{equation}
\label{index-1} \nu_{\pm} = \sqrt{\tilde{\xi}^2 + \eta^2} \pm \frac{1}{2} \epsilon (\eta).
\end{equation}
Substituting these solutions with the relation (\ref{def-f}) into the spinor $\psi$ in (\ref{solution-scalar-1}) and using the Dirac equation (\ref{dirac-scalar-1}) together with the recurrence
relations of the Bessel equations one can also derive the following relations between coefficients
\begin{equation}
\label{relation-coefficient-1} A_{2 m} = i \epsilon(\eta E) A_{1 m},   \qquad B_{2 m} = i \epsilon(\eta E) B_{1 m}.
\end{equation}
The eigenspinor $\chi(\rho)$ for the radial equation (\ref{radial-1}) is then obtained to be
\begin{eqnarray}
\label{final-1}
\left( \begin{array}{c} \dst \chi_{1m}(\rho) \\ \chi_{2m}(\rho) \end{array}\right)
 = U_c^{\dagger} \left( \begin{array}{c}
                    \dst A_{1 m} H_{\nu_-}^{(1)} (\rho) + B_{1 m} H_{\nu_-}^{(2)} (\rho)\\
                    \dst i \epsilon(\eta E) A_{1 m} H_{\nu_+}^{(1)} (\rho) + i \epsilon(\eta E) B_{1 m} H_{\nu_+}^{(2)} (\rho)
                   \end{array}\right).
\end{eqnarray}

To determine the coefficients $A_{1m}$ and $B_{1m}$ for each $m$ we use the boundary conditions given in Eq.~(\ref{boundary}). For the components of the spinor solution
(\ref{solution-scalar-1}) the boundary conditions give
\begin{eqnarray}
\label{boundary-conditions-2} \eqalign{\chi_{2 m} (\rho_1 / |\tilde{E}|) = -i \tau \chi_{1 m} (\rho_1 / |\tilde{E}|),\cr
 \chi_{2 m} (\rho_2 / |\tilde{E}|) = i \tau \chi_{1 m}(\rho_2 / |\tilde{E}|),}
\end{eqnarray}
where $\rho_1 = |\tilde{E}| (a - w/2)$ and $\rho_2 = |\tilde{E}| (a + w/2)$. Defining
\begin{eqnarray}
\label{con-def-1} \eqalign{Y^{(i)} (\nu_1, \nu_2; \rho) = \tan \frac{\phi}{2} H_{\nu_1}^{(i)} (\rho) - \epsilon (\eta E) H_{\nu_2}^{(i)} (\rho)   \cr Z^{(i)} (\nu_1, \nu_2; \rho) =
H_{\nu_1}^{(i)} (\rho) + \epsilon (\eta E) \tan \frac{\phi}{2} H_{\nu_2}^{(i)} (\rho)}
\end{eqnarray}
with $i = 1$ or $2$, the boundary conditions (\ref{boundary-conditions-2}) read
\begin{eqnarray}
\label{boundary-conditions-3} \eqalign{\left[ Y^{(1)} (\nu_-, \nu_+; \rho_2) + \tau Z^{(1)} (\nu_-, \nu_+; \rho_2) \right]A_{1m}  \cr +  \left[ Y^{(2)} (\nu_-, \nu_+; \rho_2) +
\tau Z^{(2)} (\nu_-, \nu_+; \rho_2) \right]B_{1m} = 0 \cr  \left[ Y^{(1)} (\nu_-, \nu_+; \rho_1) - \tau Z^{(1)} (\nu_-, \nu_+; \rho_1) \right]A_{1m}  \cr  +  \left[ Y^{(2)}
(\nu_-, \nu_+; \rho_1) - \tau Z^{(2)} (\nu_-, \nu_+; \rho_1) \right]B_{1m} = 0.}
\end{eqnarray}
The secular equation requires then the following relation
%\begin{widetext}
\begin{equation}
\label{scalar-spectrum-cond-1} \fl\frac{Y^{(1)} (\nu_-, \nu_+; \rho_2) + \tau Z^{(1)} (\nu_-, \nu_+; \rho_2)}
                {Y^{(1)} (\nu_-, \nu_+; \rho_1) - \tau Z^{(1)} (\nu_-, \nu_+; \rho_1)}
         =  \frac{Y^{(2)} (\nu_-, \nu_+; \rho_2) + \tau Z^{(2)} (\nu_-, \nu_+; \rho_2)}
                {Y^{(2)} (\nu_-, \nu_+; \rho_1) - \tau Z^{(2)} (\nu_-, \nu_+; \rho_1)}.\quad
\end{equation}
%\end{widetext}
From this the spectrum of energy eigenvalues of a Dirac electron in the graphene ring can be calculated. It should be noted here that the scalar coupling is implied in the orders
$\nu_{\pm}$ of the Hankel functions (see Eq.~(\ref{index-1})) and hence the  Eq.~(\ref{scalar-spectrum-cond-1}) with $\tilde{\xi} = 0$ is identical to the energy eigenvalue
equation derived in Ref.~\cite{recher07}.

To obtain an explicit expression of the eigenvalue spectrum we assume
\begin{equation}
\label{analytical-scalar-1} \frac{w}{2 a} \sim \frac{v_F}{ |E| a} << 1
\end{equation}
and exploit the asymptotic formula of the Hankel functions for large $\rho$ in the condition (\ref{scalar-spectrum-cond-1}). This gives the following equation
\begin{equation}
\label{analytical-scalar-2} |E| = \varepsilon_n + \frac{v_F}{w} \epsilon (\eta) \sqrt{\eta^2 + \tilde{\xi}^2} \left(\frac{v_F}{ |E| a}\right) \Omega_{\tau,m,s}
(\alpha;\tilde{\xi}; E)
\end{equation}
where $\varepsilon_n = v_F(n + 1/2)/w$ $(n=0,1,2,\dots)$, and
\begin{eqnarray}
\label{analytical-scalar-3} \fl\eqalign{\Omega_{\tau,m,s} (\alpha;\tilde{\xi}; E) = \epsilon (\omega_1) \tau \sqrt{\omega_1^2 - \omega_2}, \cr \omega_1 = \tau \epsilon(\eta)
\left( \frac{w}{2 a}\right)  \sqrt{\eta^2 + \tilde{\xi}^2} - \epsilon (E)\left(\frac{v_F}{2 |E| a}\right)
                         \frac{|\eta|}{\sqrt{\eta^2 + \tilde{\xi}^2}} + \epsilon (\eta) \frac{\tilde{\xi}}{\sqrt{\tilde{\xi}^2 + \eta^2}},
\cr \omega_2 = \left(\frac{v_F}{ |E| a}\right) \left( \frac{\tilde{\xi}^2}{\tilde{\xi}^2 + \eta^2} \right) \left[ \left(\frac{3v_F}{4|E| a}\right) \tilde{\xi}^2 + \tau \epsilon
(E) \left( \frac{w}{a} \right) \eta \right].}
\end{eqnarray}
Solving Eq.~(\ref{analytical-scalar-2}) by iteration and keeping leading terms, the energy eigenvalues are obtained to be
\begin{equation}
\label{analytical-scalar-4} E_{nm}(\tau,\xi,\alpha) = \pm \varepsilon_n \pm \mcl{E}_{nm}(\tau,\xi,\alpha),
\end{equation}
where
\begin{equation}
\label{analytical-scalar-5} \fl\dst \mcl{E}_{nm}(\tau,\xi,\alpha) = \left( \frac{2 a}{w} \right) \left[\frac{1}{2\varepsilon_n}\left(\frac{v_F}{a}\right)^2\right] \epsilon(\eta)
\sqrt{\eta^2 + \xi^2} \Omega_{\tau,m,s} (\alpha;\xi; \pm \varepsilon_n).
\end{equation}

%%%%%%%%%%%%%%%%%%%%%%%%%%%%%%%%%%%%%%%%%%%%%%%%%%%%%%%%%%%%%%%%
\begin{figure}[t]
\begin{center}
\includegraphics[height=5.0cm]{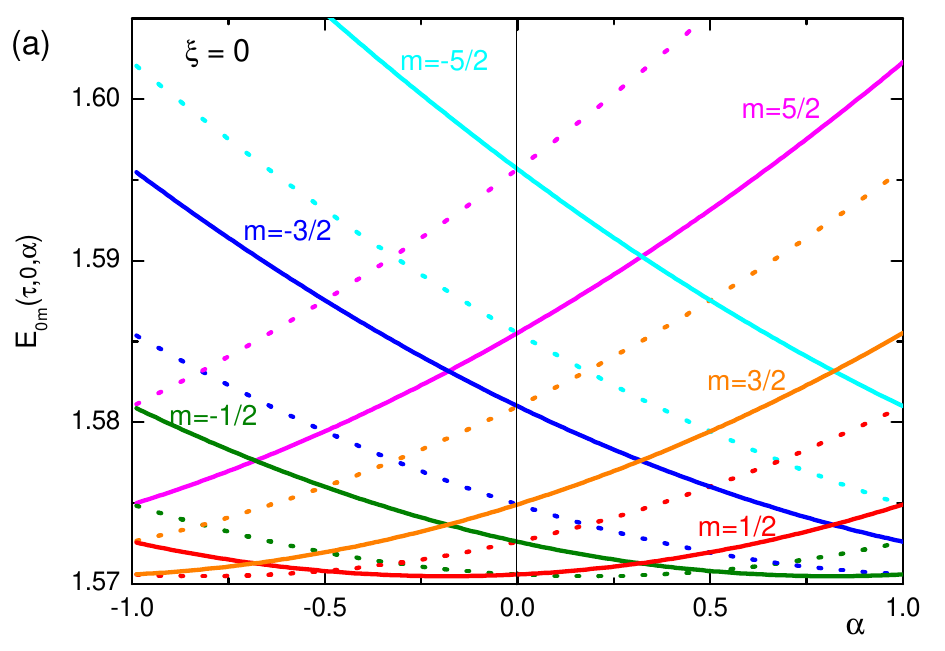}
\hspace{0.3in}
\includegraphics[height=5.2cm]{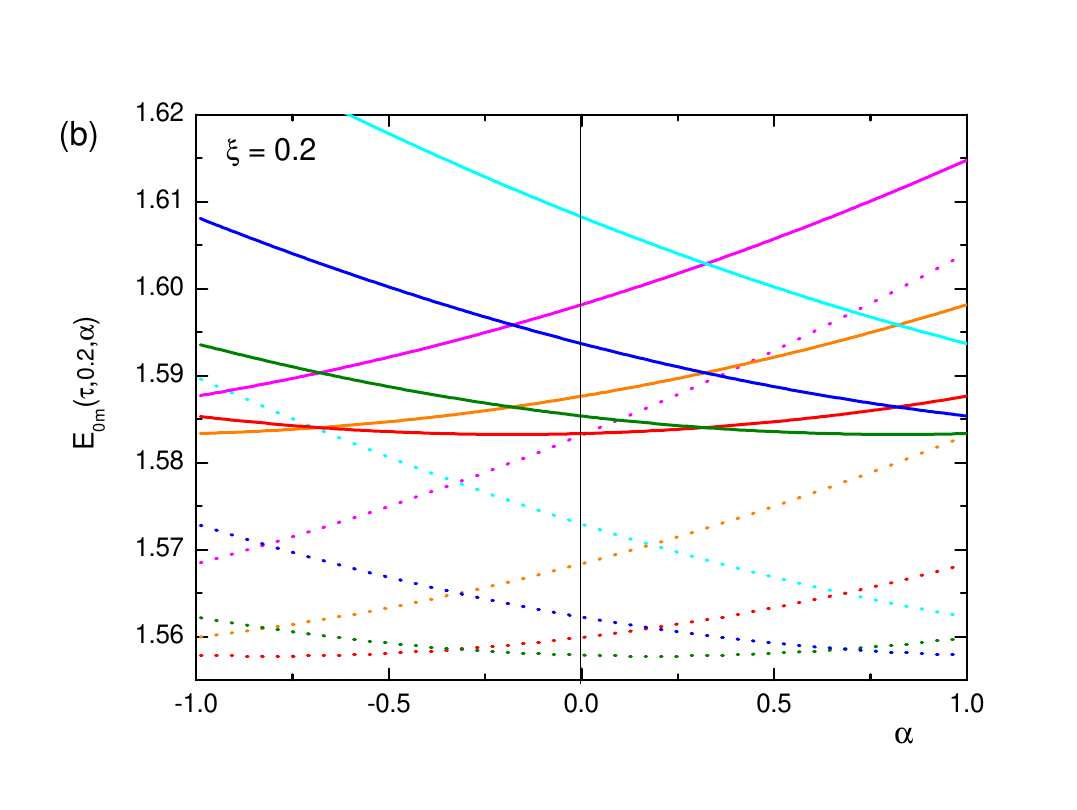}
\includegraphics[height=5.2cm]{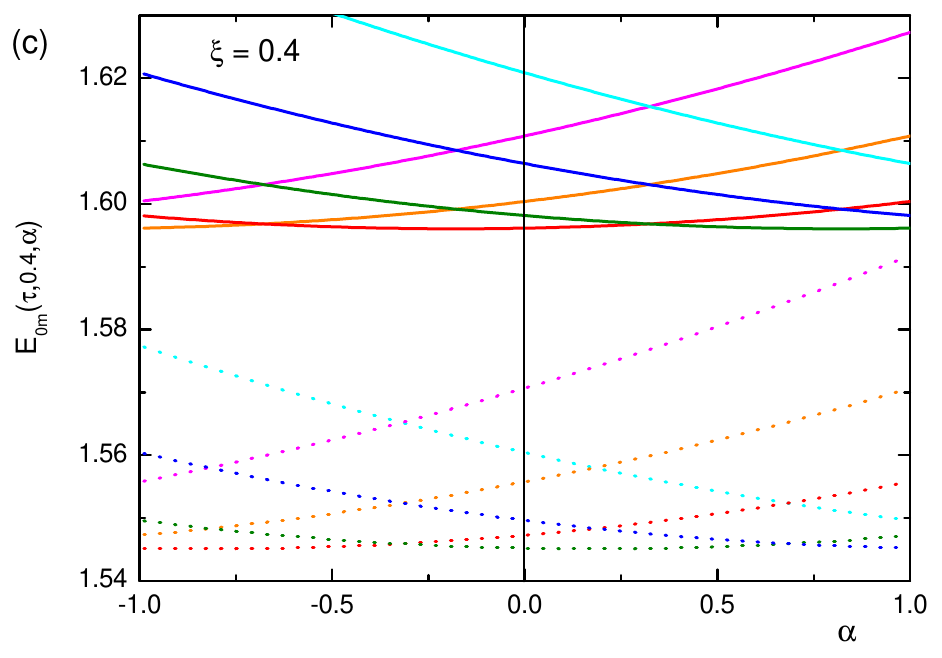}
\caption{\label{fig2}Energy spectrum of positive $E_{0m}(\tau,\xi,\alpha)$. For convenience we choose $v_F = w = 1$ and $a = 10$. The solid and dotted lines correspond to $\tau =1$ and $\tau = -1$,
respectively. The colors pink, orange, red, green, blue, and cyan correspond to $m=5/2$, $3/2$, $1/2$, $-1/2$, $-3/2$, and $-5/2$, respectively. (a) is the energy spectrum when $\xi = 0$, the case
without interaction which is identical to Ref.~\cite{recher07}. (b) and (c) are the energy spectra when the interaction exists. Note that, besides the breaking of valley degeneracy, the
spectrum of the $\tau = +1$ valley is raised and the spectrum of the $\tau = -1$ valley is lowered.}
\end{center}
\end{figure}
%%%%%%%%%%%%%%%%%%%%%%%%%%%%%%%%%%%%%%%%%%%%%%%%%%%%%%%%%%%%%%%%%%

In Fig.~\ref{fig2} we plot the first few positive energy levels with $n=0$ as a function of the magnetic flux $\alpha$ for different values of $\xi$. In the figure the solid and dotted lines
correspond to $\tau = 1$ and $\tau = -1$, respectively. Fig.~\ref{fig2}(a) is the plot of $\alpha$-dependence when the Coulomb interaction is zero, which is the same situation as
Ref.~\cite{recher07}. For the analysis of the energy spectrum with $\xi=0$, we let $E_{0m}(\tau,0,\alpha) = E(\tau, m, \alpha)$. When $\alpha = 0$, since the time reversal symmetry by $\mcl{T}$ is
preserved,  there are degeneracies between $E(1, m,0)$ and $E(-1, -m,0)$. However, the intravalley degeneracy is broken, that is, $E(\tau, m,0) \neq E(\tau,-m,0)$ because of the confinement potential
that breaks the effective time reversal symmetry by $\bs{\Theta}$. Obviously, when $\alpha \neq 0$, these degeneracies are broken due to the AB potential.

Fig.~\ref{fig2}(b) and Fig.~\ref{fig2}(c) are the plots of $\alpha$-dependence when $\xi = 0.2$ and $\xi=0.4$, the case of repulsive interaction. As noted from the figures, the
splitting of the intravalley degeneracy still exists due to the mass terms of the scalar coupling and the confinement potential. What is more interesting here is that the scalar
coupling raises the whole energy spectrum of the $\tau = +1$ valley and lowers the whole energy spectrum of the $\tau=-1$ valley. The situation will be reversed if the interaction
is attractive; the $\tau=-1$ valley is raised, while the $\tau=+1$ valley is lowered. As explained in Sec.~\ref{sec2} this effect is ascribed to the TRS breaking by the scalar
coupling of $\xi/r$ entered as a mass term in the Dirac equation.

Another interesting feature to be noted is that the separation between the two spectra gets larger as the interaction strength increases. To see the energy difference between the
two separated spectra, we look into the $\tau$-dependent part in the Eq. (\ref{analytical-scalar-5}) for a fixed set of $(n,m,\alpha)$. The energy difference is then determined by
the term $\mcl{E}_{nm}(\tau,\xi,\alpha)$ which can be positive or negative depending on the sign of $\tau$ together with $\eta$ and $\omega_1$. With the assumption
(\ref{analytical-scalar-1}), since $\eta \omega_1 \sim \vartheta\xi$ ($\vartheta$ a positive constant), we may write $\mcl{E}_{nm}(\tau,\xi,\alpha) \propto \epsilon(\tau\xi)
\xi\sqrt{\eta^2 +\xi^2}$, which becomes positive when $\tau \xi
>0$ and negative when $\tau \xi <0$. Thus, for a repulsive interaction for which $\xi>0$, the energy spectrum of $\tau=+1$ valley is raised, while that of $\tau = -1$ valley is lowered; for
the attractive interaction, since $\xi < 0$, the opposite occurs. The energy difference between the two spectra is then $\Delta \mcl{E}_{nm} \propto \xi\sqrt{\eta^2 +\xi^2}$, which explains the
increase of the separation with $\xi$.

%%%%%%%%%%%%%%%%%%%%%%%%%%%%%%%%%%%%%%%%%%%%%%%%%%%%%%%%%
\begin{figure}[t]
\begin{center}
\includegraphics[height=5cm]{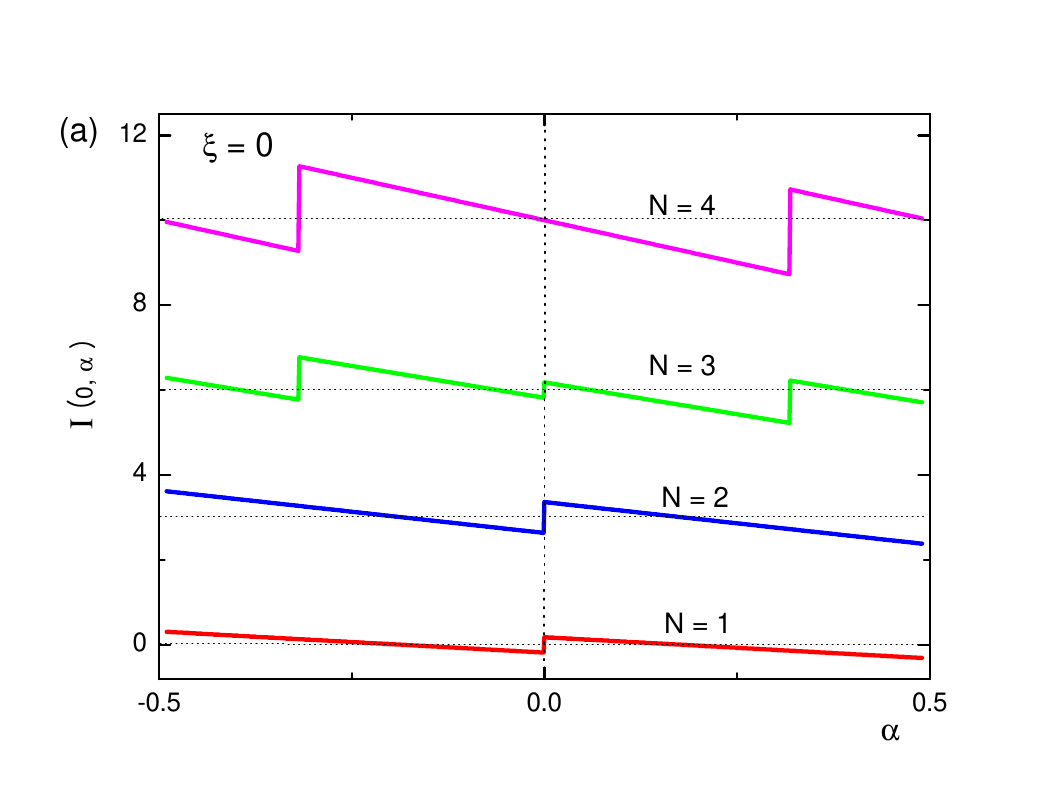}
\hspace{0.3in}
\includegraphics[height=5cm]{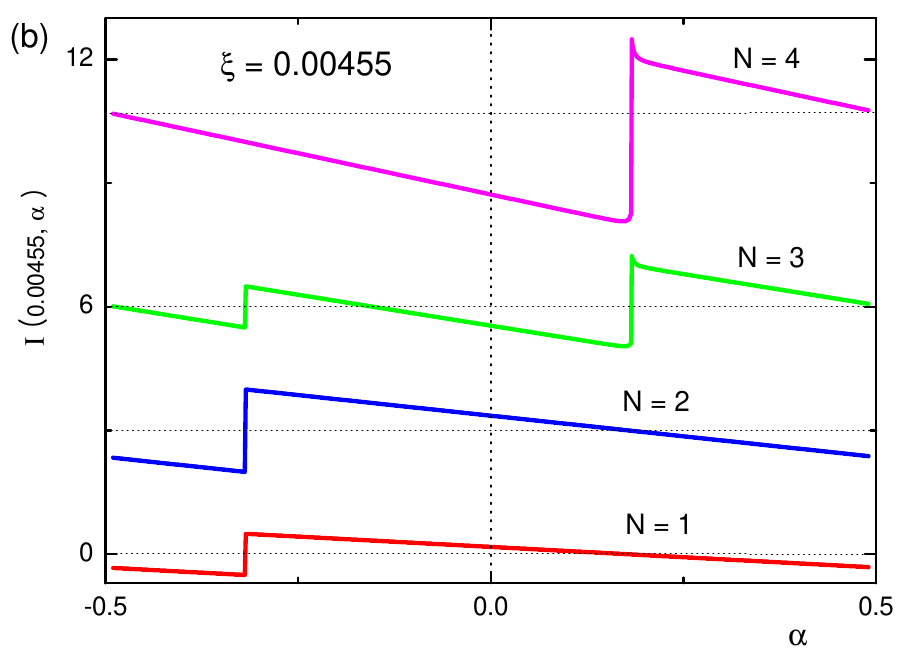}
\includegraphics[height=5cm]{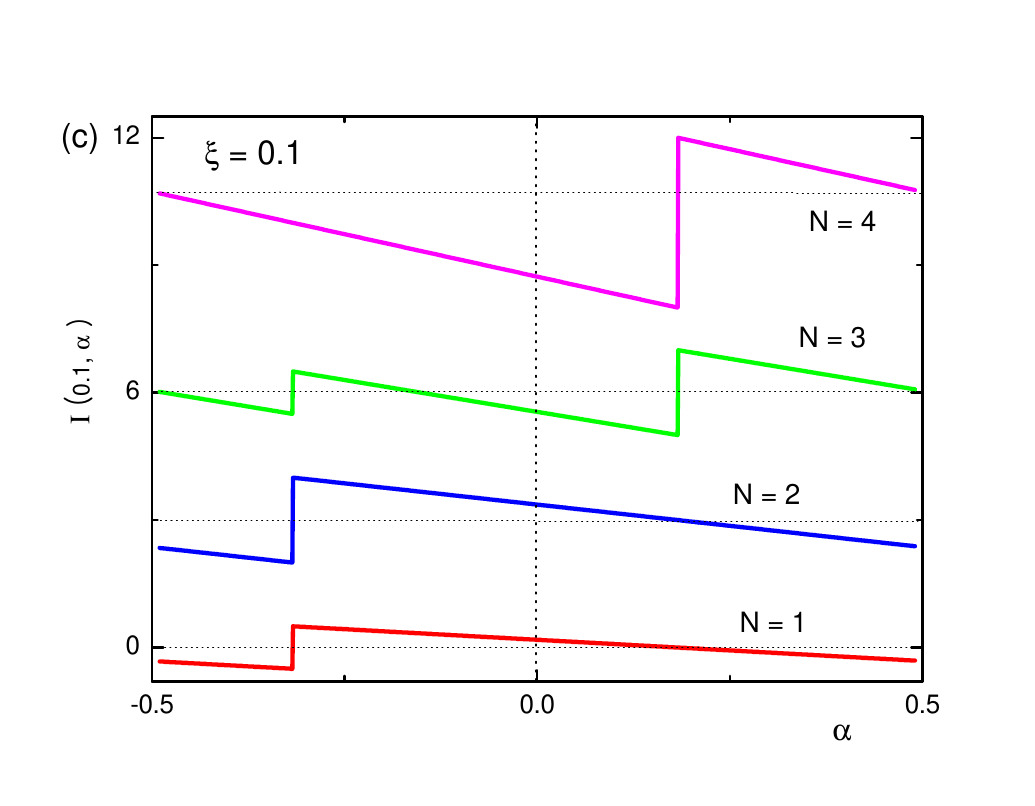}
\caption{\label{fig3}Persistent currents as a function of $\alpha$ for different values of $\xi$. $N$ is the number of electrons in the occupied states. (a) is when $\xi =0$, the case without Coulomb
interaction. (b) and (c) are the cases with Coulomb interaction through the scalar coupling. Note that persistent currents are symmetric about $\alpha=0$ when $\xi=0$, but asymmetric about $\alpha=0$
when $\xi>0$. The latter is ascribed to the separation between the spectrum of $\tau=+1$ valley and that of $\tau=-1$ valley.}
\end{center}
\end{figure}
%%%%%%%%%%%%%%%%%%%%%%%%%%%%%%%%%%%%%%%%%%%%%%%%%%%%%%%%%%%

A possible way of observing the splittings of valley degeneracies is to measure the magnetic moment induced by persistent current in the graphene ring \cite{lddb}. The persistent current can be
obtained from the energy eigenvalues and, at zero temperature, it is given by
\begin{eqnarray}
\dst I(\xi,\alpha) = -\sum_{\tau=\pm1}\sum_n\sum_m\frac{\partial }{\partial \alpha}E_{nm}(\tau,\xi,\alpha).
\end{eqnarray}
In Fig.~\ref{fig3} we present the persistent current as a function of the magnetic flux $\alpha$ for different values of $\xi$ and the number of electrons $N$ (including spin).
Fig.~\ref{fig3}(a) is the plot when $\xi=0$, the case without interaction. Since the intervalley degeneracy exists the current is symmetric about $\alpha=0$ when there are equal
number of electrons at each valley, as can be seen from the figure. When $\xi > 0$ and large, the separation between the two spectra of $\tau=\pm1$ valleys is large. Since the
electrons occupy from the lowest levels at zero temperature the states of the lower spectrum (the $\tau=-1$ valley) will be occupied first, while those of the higher spectrum (the
$\tau=+1$ valley) are almost empty. Thus, the persistent current is contributed largely by the lower valley electrons, and becomes essentially a single valley phenomenon, known as
valley polarization. This should produce an asymmetric persistent current about $\alpha = 0$ and a finite value at $\alpha=0$, which are illustrated in Figs.~\ref{fig3}(b) and
\ref{fig3}(c). As we can notice in Fig.~\ref{fig3}(b), even for very small value of the interaction strength, the qualitative feature is quite different from the case without the
interaction because of its role of symmetry breaking. We also emphasize that the $\alpha$-dependences of persistent currents for $\xi > 0$ are essentially the same because the
electrons occupy only the energy levels in the lower spectrum corresponding to the $\tau=-1$ valley.

\section{ABC problem with vector coupling}\label{sec4}

In this section we consider the vector coupling of the potential $\xi/r$. Here we will use the term Coulomb potential because the vector coupling is related to the electric
Coulomb potential.  Using the Hamitonian given in Eq.~(\ref{hvector}) the Dirac equation inside the ring can be written as
\begin{eqnarray}\label{dirac-vector-1}
\dst {\bs\sigma} \cdot \left(\bs{p} + e\bs{A}\right)\psi = \left(\tilde{E} - \frac{\tilde{\xi}}{r}\right) \psi
\end{eqnarray}
 Operating ${\bs\sigma} \cdot \left(\bs{p} + e\bs{A}\right)$ on the equation and following the same procedure as in the previous section the reduced radial equation is given by
\begin{eqnarray}
\label{vector-radial-2} \eqalign{\left[ \partial_r^2 + \frac{1}{r} \partial_r - \frac{\eta^2 + 1/4}{r^2} + \left( \tilde{E}^2 - \frac{\tilde{\xi}}{r} \right)^2 \right]
\left( \begin{array}{c} f_{1m}(r) \\ f_{2m}(r) \end{array}\right)\cr = -\frac{1}{r^2} \epsilon(\eta) \sqrt{\eta^2 - {\tilde{\xi}}^2} \sigma_3 \left( \begin{array}{c} f_{1m}(r) \\ f_{2m}(r) \end{array}\right).}
\end{eqnarray}
Here, the two components $(f_{1m}(r), f_{2m}(r))$ are related to the spinor $\chi(r)$ as follows:
\begin{equation}
\label{vector-def-f} \dst \chi(r) = \left( \begin{array}{c} \chi_{1m}(r) \\ \chi_{2m}(r) \end{array}\right) = U_v^{-1} \left( \begin{array}{c} f_{1m}(r) \\ f_{2m}(r) \end{array}\right),
\end{equation}
where the transformation matrix $U_v$ is defined by
\begin{eqnarray}
\label{diagonal-2}
\eqalign{U_v^{\pm1} = \cosh\frac{\phi}{2}\, \sigma_0 \pm \epsilon(\tilde{\xi})\epsilon(\eta)\,\sinh\frac{\phi}{2}\,\sigma_2\cr
\tanh\frac{\phi}{2} = \epsilon(\tilde{\xi})\frac{|\eta| - \sqrt{\eta^2 - {\tilde{\xi}}^2}}{\tilde{\xi}}}
\end{eqnarray}
and satisfies the matrix diagonalization $\eta \sigma_3 - i \tilde{\xi} \sigma_1 = \epsilon(\eta) \sqrt{\eta^2 - \tilde{\xi^2}} U_v^{-1} \sigma_3 U_v$. Note that $U_v$ in Eq.~(\ref{vector-def-f}) is
not unitary.

The solutions of Eq.~(\ref{vector-radial-2}) are expressed in terms of the confluent hypergeometric functions:
\begin{eqnarray}
\label{vector-solution-1} \fl \eqalign{f_{1 m}(\rho) = e^{i\rho}(-2i\rho)^{t_-}  \left[ C_{1m} M(a_-,\, b_-,\, - 2 i \rho) + D_{1m} U(a_-,\, b_-,\, - 2 i \rho)\right]     \cr
f_{2 m}(\rho) = e^{i\rho}(-2i\rho)^{t_+}  \left[ C_{2m} M(a_+,\, b_+,\, - 2 i \rho) + D_{2m} U(a_+,\, b_+,\, - 2 i \rho)\right]}
\end{eqnarray}
where $M(a,b,\rho)$ and $U(a,b,\rho)$ are the Kummer's functions~\cite{abromowitz}, $\rho=|\tilde{E}|r$, and  $a_{\pm}$, $b_{\pm}$ and $t_{\pm}$ are given by
\begin{eqnarray}
\label{define-t-k} a_{\pm} &=& t_{\pm} +\frac{1}{2}+i\epsilon(\tilde{E})\tilde{\xi},\quad b_{\pm} =2t_{\pm} +1\nonumber\\
t_{\pm} &=& \sqrt{\eta^2 -{\tilde{\xi}}^2} \pm \frac{\epsilon(\eta)}{2}.
\end{eqnarray}
By the same method as in Sec.~II and exploiting the recurrence relations and differential properties of the confluent hypergeometric functions, one can derive the following relations between
coefficients after some algebra;
\begin{eqnarray}
\label{vector-relation-coefficient-1} C_{2m} = \epsilon(E) c_{21} C_{1m} \qquad D_{2m} = \epsilon(E) d_{21} D_{1m} ,
\end{eqnarray}
where
\begin{eqnarray}
\label{vector-relation-coefficient-2} \fl c_{21} = - \left( \frac{|\eta|}{4 (\eta^2 - \tilde{\xi}^2) +2 \sqrt{\eta^2 - \tilde{\xi}^2}} \right)^{\epsilon(\eta)},
\qquad d_{21} = \left( \frac{\sqrt{\eta^2 - \tilde{\xi}^2} + i \epsilon(E) \tilde{\xi}}{|\eta|} \right)^{\epsilon(\eta)} .
\end{eqnarray}
Substituting the solutions (\ref{vector-solution-1}) with the relation (\ref{vector-relation-coefficient-1}) into the transformation equation (\ref{vector-def-f}) we can write the eigenspinor
$\chi(\rho)$ as
\begin{eqnarray}
\label{vector-solution-4}
\left( \begin{array}{c} \chi_{1m}(\rho) \\ \chi_{2m}(\rho) \end{array}\right) = e^{i\rho}(-2i\rho)^{t_-} \left( \begin{array}{cc}  m_1(\rho) & u_1(\rho) \\ m_2(\rho) & u_2(\rho) \end{array}\right)
\left( \begin{array}{c} C_{1m} \\ D_{1m} \end{array}\right)
\end{eqnarray}
where
%\begin{widetext}
\begin{eqnarray}
\label{vector-solution-5} \fl\eqalign{m_1 (\rho) = \cosh\frac{\phi}{2}\, M(a_-,b_-,-2i\rho) + i \epsilon(\tilde{\xi}\eta E)(-2i\rho)^{\epsilon(\eta)}\,c_{21}\sinh\frac{\phi}{2}
M(a_+,b_+,-2i\rho)    \cr
u_1 (\rho) = \cosh\frac{\phi}{2}\, U(a_-,b_-,-2i\rho) + i \epsilon(\tilde{\xi}\eta E)(-2i\rho)^{\epsilon(\eta)}\,d_{21}\sinh\frac{\phi}{2}
U(a_+,b_+,-2i\rho)    \cr
m_2 (\rho)  =  \epsilon (E) (-2i\rho)^{\epsilon(\eta)}\,c_{21}\cosh\frac{\phi}{2}\, M(a_+,b_+,-2i\rho) - i \epsilon(\tilde{\xi}\eta)\sinh\frac{\phi}{2} M(a_-,b_-,-2i\rho)    \cr
u_2 (\rho)  =  \epsilon (E) (-2i\rho)^{\epsilon(\eta)}\,d_{21}\cosh\frac{\phi}{2}\, U(a_+,b_+,-2i\rho) - i \epsilon(\tilde{\xi}\eta)\sinh\frac{\phi}{2} U(a_-,b_-,-2i\rho).}
\end{eqnarray}
%\end{widetext}

To proceed we use the same infinite mass boundary condition introduced in Sec.~II. Inserting Eq.~(\ref{vector-solution-4}) into the boundary condition (\ref{boundary-conditions-2}) it is
straightforward to show
\begin{eqnarray}
\label{vector-boundary-condition-4} \eqalign{\left[ m_2 (\rho_2) - i \tau m_1 (\rho_2) \right]C_{1m} + \left[ u_2 (\rho_2) - i \tau u_1 (\rho_2) \right]D_{1m} = 0   \cr
 \left[ m_2 (\rho_1) + i \tau m_1 (\rho_1) \right]C_{1m} + \left[ u_2 (\rho_1) + i \tau u_1 (\rho_1) \right]D_{1m} = 0,}
\end{eqnarray}
where $\rho_1 =|\tilde{E}|(a-w/2)$ and $\rho_2 =|\tilde{E}|(a+w/2)$. The secular equation requires then the following condition:
\begin{equation}
\label{vector-spectrum-condi-2} \frac{m_2 (\rho_2) - i \tau m_1 (\rho_2)}{m_2 (\rho_1) + i \tau m_1 (\rho_1)}  = \frac{u_2 (\rho_2) - i \tau u_1 (\rho_2)}{u_2 (\rho_1) + i \tau u_1 (\rho_1)}.
\end{equation}
%%%%%%%%%%%%%%%%%%%%%%%%%%%%%%%%%%%%%%%%%%%%%%%%%%%%%%%%%
\begin{figure}[ht!]
\begin{center}
\includegraphics[height=5cm]{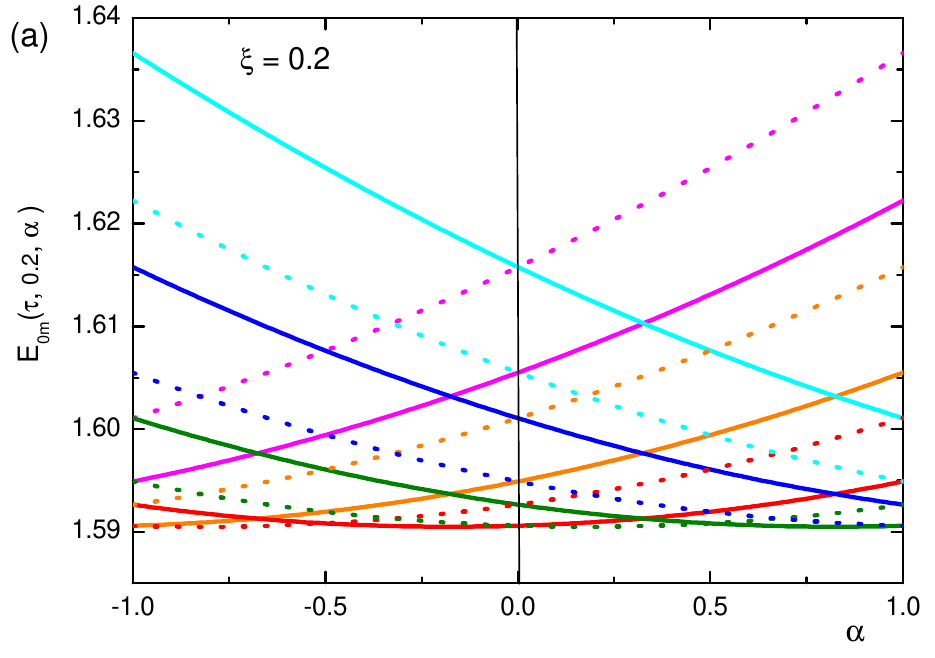}
\hspace{0.3in}
\includegraphics[height=5cm]{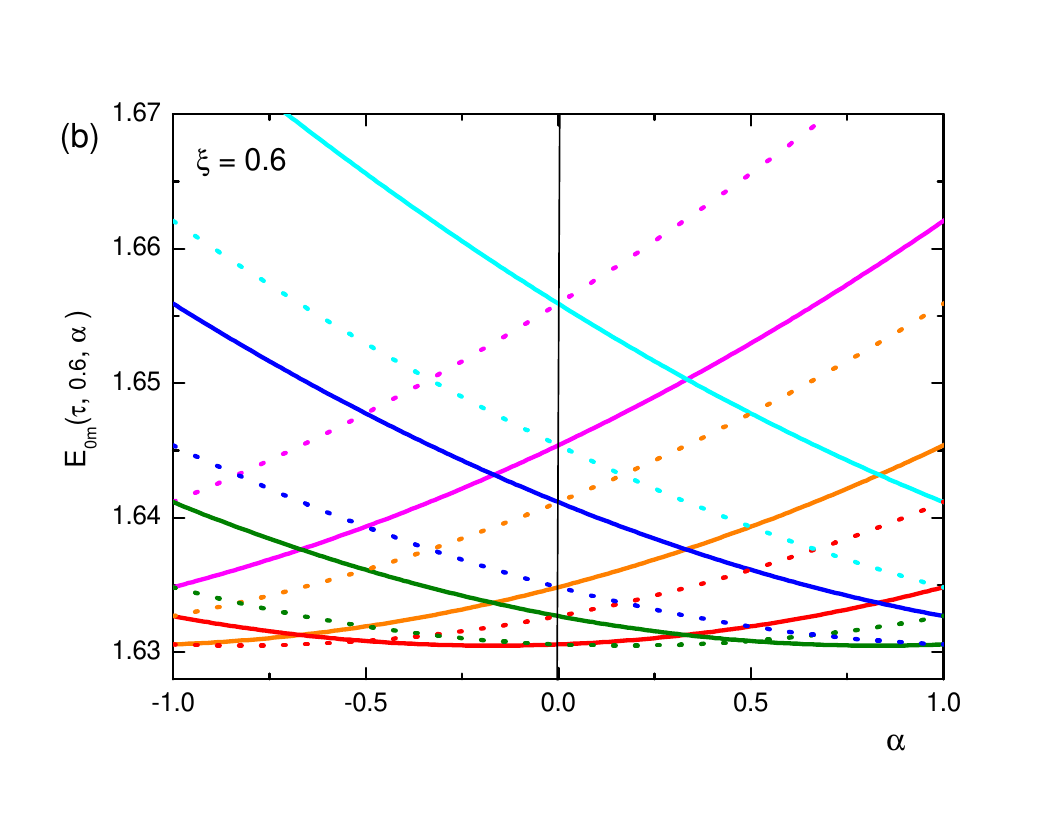}
\caption{\label{fig4}Energy spectrum for the vector coupling. For convenience we choose $v_F = w = 1$ and $a = 10$. The solid and dotted lines correspond to $\tau =1$ and $\tau = -1$, respectively.
The colors pink, orange, red, green, blue, and cyan correspond to $m=5/2$, $3/2$, $1/2$, $-1/2$, $-3/2$, and $-5/2$, respectively. (a) and (b) are energy spectra for $\xi = 0.2$ and $\xi = 0.6$,
respectively. Note that the intervalley symmetry about $\alpha=0$ is still kept here even in the presence of the  potential and the whole spectrum is raised by $\xi/a$ compared to the
Fig.~\ref{fig2}(a).}
\end{center}
\end{figure}
%%%%%%%%%%%%%%%%%%%%%%%%%%%%%%%%%%%%%%%%%%%%%%%%%%%%%%%%%%
To obtain the eigenvalue spectrum we impose the same condition with Eq.~(\ref{analytical-scalar-1}) and use the asymptotic formula \footnote{For large $|z|$
\begin{eqnarray*}
\label{analytical-vector-1} \fl M(a, b, z) \sim e^z z^{a-b} \frac{\Gamma (b)}{\Gamma (a)} \sum_{n=0} \frac{(b-a)_n (1-a)_n}{n!} z^{-n},\quad  U(a, b, z) \sim z^{-a} \sum_{n=0}
\frac{(a)_n 1 + a - b)_n}{n!} (-z)^{-n},
\end{eqnarray*}
where $(a)_n = a (a + 1) \cdots (a + n - 1)$; see Ref. \cite{abromowitz}} of the confluent hypergeometric functions for large $\rho$. Expanding Eq.~(\ref{vector-spectrum-condi-2})
up to the second order of perturbation, one can show that the energy eigenvalue satisfies
%\begin{widetext}
\begin{eqnarray}
\label{analytical-vector-2} \fl |E| = \frac{v_F}{w} \left[ \left(n+ \frac{1}{2}\right) \pi + 2 \epsilon (E) \tilde{\xi} \left(\frac{w}{2 a}\right) + \eta^2 \left( \frac{v_F}{|E| a}
\right) \left(\frac{w}{2 a}\right)  - \frac{\tau}{2} \epsilon (E) \eta \left( \frac{v_F}{|E| a} \right)^2 \right],
\end{eqnarray}
%\end{widetext}
where $n$ is a non-negative integer. Solving Eq.~(\ref{analytical-vector-2}) by iteration and keeping only leading terms, we obtain the energy eigenvalues
\begin{equation}
\label{analytical-vector-3} \dst E_{n m}(\tau,\xi,\alpha) = \pm \varepsilon_n  \pm \lambda_n \eta \left( \eta \mp \frac{\tau}{\left(n+\frac{1}{2}\right) \pi} \right) +
\frac{\xi}{a},
\end{equation}
where $\varepsilon_n = v_F(n + 1/2)/w$ $(n=0,1,2,\dots)$ and $\lambda_n = (v_F/a)^2/2\varepsilon_n$, and the $\alpha$ dependence is through the variable $\eta$ defined in Eq.~(\ref{eta-def-1}).

We can immediately see from Eq.~(\ref{analytical-vector-3}) that the effect of the potential $\xi/r$ in the vector coupling  is merely an additive constant $(\xi/a)$, shifting the
whole spectrum of energy eigenvalues. It also reveals the complete decoupling between the AB and the Coulomb effects, because the Coulomb effect appears in the first order whereas
the AB effect arises in the second order perturbation.  In Fig.~\ref{fig4}, we show the energy spectrum in which the symmetry between the two valley spectra about $\alpha = 0$ is
manifested, representing the existence of the intervalley degeneracy. As discussed in Sec.~\ref{sec2}, the Hamiltonian without AB potential in Eq.~(\ref{dirac-vector-1}) is
invariant under the time reversal operators $\mcl{T}$ and $\bs{\Theta}$, and hence the vector coupling breaks neither the intervalley nor the intravalley degeneracies. It is thus
expected that the eigenvalue spectrum in the vector coupling has essentially the same feature as the spectrum without the Coulomb interaction. By the same reason, the Coulomb
potential in the vector coupling does not alter the behavior of the persistent current and hence it is symmetric about $\alpha=0$ as shown in Fig.~\ref{fig3}(a).

%%%%%%%%%%%%%%%%%%%%%%%%%%%%%%%%%%%%%%%%%%%%%%%%%%%%%%%%%
\begin{figure}[ht!]
\begin{center}
\includegraphics[height=5cm]{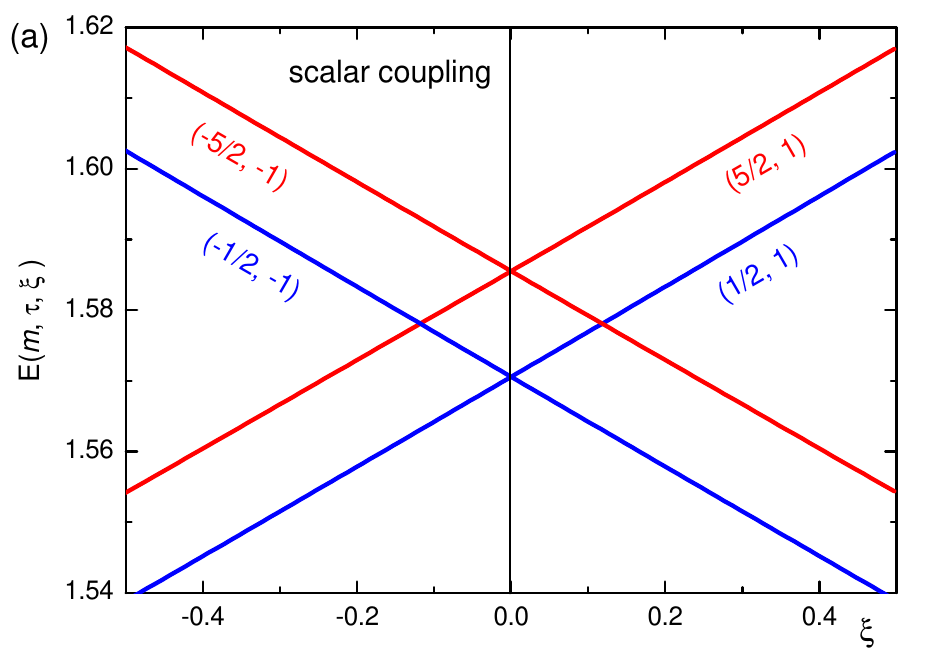}
\hspace{0.3in}
\includegraphics[height=5cm]{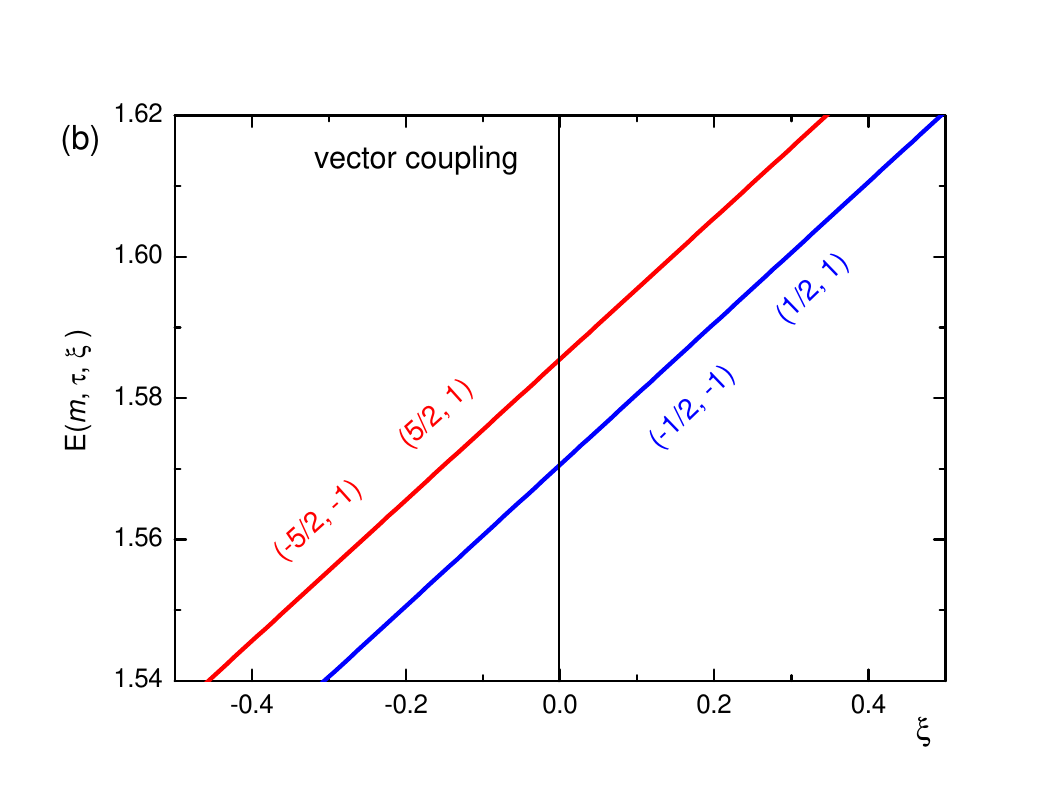}
\caption{\label{fig5}The $\xi$-dependence of the energy spectrum when there is no AB potential (that is, when $\alpha=0$). For convenience we let $E_{0m}(\tau,\xi,0) = E(m,\tau,\xi)$, and the
parenthesis denotes $(m, \tau)$. (a) and (b) correspond to the scalar coupling and the vector coupling for $m=\pm 1/2, \pm 5/2$ and $\tau = \pm 1$, respectively. The splitting of intervalley
degeneracy is clearly seen in (a), but it is not broken in (b).}
\end{center}
\end{figure}
%%%%%%%%%%%%%%%%%%%%%%%%%%%%%%%%%%%%%%%%%%%%%%%%%%%%%%%%%%%

In Fig.~\ref{fig5}, we also present the energy spectrum as a function of $\xi$ without the AB potential to compare the bare effects of the Coulomb type potential between the scalar coupling and the vector
coupling. Fig.~\ref{fig5}(a) shows the splitting of intervalley degeneracy due to the symmetry-breaking term in the scalar coupling. However, in Fig.~\ref{fig5}(b), we see that no intervalley
splitting exists in the case of vector coupling. This also confirms that, in the case of vector coupling, the Coulomb potential alone does not break the intervalley degeneracy, as explained by the
consideration of the time reversal symmetry.

\section{Conclusion}\label{sec5}
In this paper we have considered the ABC problem in a graphene ring with magnetic flux tube and a Coulomb type potential $\xi/r$ at its center. We have investigated the effects of
the potential on the energy spectrum in two different ways: the scalar coupling and the vector coupling. In the scalar coupling the potential enters the 2D Dirac equation as a
symmetry-breaking mass term, so that both the intervalley and the intravalley degeneracies are broken. The main effect of the interaction appears as the separation between the
energy spectrum of one valley and that of the other valley, which is attributed to the breaking of the intervalley degeneracy; for the repulsive interaction the
$\tau=+1$ ($\bs{K}$) valley is lifted while the $\tau=-1$ ($\bs{K'}$)  valley is lowered, and vice versa for the attractive interaction. Contrary to the scalar coupling, the
potential in the vector coupling does not break any symmetry and only shifts the AB energy spectrum, indicating essentially the same feature as the energy spectrum without a
Coulomb potential.

The results obtained here suggests that the scalar coupling of the Coulomb type potential $\xi/r$ can decouple the two valleys because of the considerable lift of energy spectrum,
so that  each valley degree of freedom becomes an independent quantity, called a valley isospin. A real experiment with graphene ring may realize this decoupling, known as the
valley polarization, which is an important element in the graphene-based electronics, called {\it valleytronics} \cite{rycerz,zhou}.

Another interesting issue related to the AB effect is when the thin magnetic flux tube is located at the center of the honeycomb lattice shown in Fig.~\ref{fig1}(a). Since the
anyon is a flux-carrying particle, a similar situation arises when there is an anyon impurity in the graphene plane. Since, in this case, the magnetic flux is located on the
graphene, we should treat the singular solution problem, which was extensively discussed in Refs.~\cite{spin-ab,cosmic} in the context of the anyonic and cosmic string theories.
Roughly speaking, there are two prescriptions for the interpretation of the singular solution: one is a mathematically-based prescription called self-adjoint extension and the
other is a physically-based prescription. We do not know yet which prescription is physically more reasonable. Probably, the real experiment with graphene may shed light on the
treatment of the singular solution. If so, our understanding on the anyonic and cosmic string theories can be greatly enhanced through the graphene physics. We will explore this
issue in the future.

{\ack Daekil Park's research was supported by the Kyungnam University Foundation Grant, 2011.}

\end{document}